\documentclass{elsart}

\usepackage{epsfig}
\usepackage{amssymb}

\begin{document}

\begin{frontmatter}

\title{Charged gravastars in higher dimensions}

\author[label1]{S. Ghosh}\ead{shnkghosh122@gmail.com},
\author[label2]{F. Rahaman}\ead{rahaman@associates.iucaa.in},
\author[label3]{B.K. Guha}\ead{bkguhaphys@gmail.com},
\author[label4]{Saibal Ray}\ead{saibal@associates.iucaa.in}

\address[label1]{Department of Physics, Indian Institute of Engineering Science and Technology, B. Garden, Howrah 711103, West Bengal, India}

\address[label2]{Department of Mathematics, Jadavpur University, Kolkata 700032, West Bengal, India }

\address[label3]{Department of Physics, Indian Institute of Engineering Science and Technology, B. Garden, Howrah 711103, West Bengal, India }

\address[label4]{Department of Physics, Government College of Engineering and Ceramic Technology, 73 A.C.B. Lane, Kolkata
	700010, West Bengal, India}

\begin{abstract}
We explore possibility to find out a new model of gravastars in
the extended $D$-dimensional Einstein-Maxwell spacetime. The class
of solutions as obtained by Mazur and Mottola of a neutral
gravastar~\cite{Mazur2001,Mazur2004} have been observed as a compitent
alternative to $D$-dimensional versions of the
Schwarzschild-Tangherlini black hole. The outer region
of the charged gravastar model therefore corresponds to a higher
dimensional Reissner-Nordstr{\"o}m black hole. In connection to
this junction conditions, therefore we have
formulated mass and the related Equation of State of the
gravastar. It has been shown that the model satisfies all the
requirements of the physical features. However, overall
observational survey of the results also provide probable
indication of non-applicability of higher dimensional approach for
construction of a gravastar with or without charge from an
ordinary $4$-dimensional seed as far as physical ground is
concerned.
\end{abstract}


\end{frontmatter}

\section{Introduction}
A decade or more ago Mazur and Mottola \cite{Mazur2001,Mazur2004} have
proposed a new solution for the endpoint of a gravitationally
collapsing neutral system. By extending the concept of Bose-Einstein condensation
to gravitational systems they constructed a cold compact object
which consists of an (i) interior de Sitter condensate phase, and (ii) exterior
Schwarzschild geometry. These are separated by a phase boundary
with a small but finite thickness $r_2-r_1=\delta$ of the thin
shell, where $r_1$ and $r_2$ represent the interior and exterior radii of the gravastar.
Therefore, the equation of state (EOS) under consideration are as follows:

I. Interior: $0 \leq r < r_1 $, ~with EOS~$ p = -\rho $,

II. Shell: $ r_1 < r < r_2 $, ~with EOS~$ p = +\rho $,

III. Exterior: $ r_2 < r $, ~with EOS~$ p = \rho =0. $

Here the presence of matter on the shell is required to achieve
the stability of the systems under expansion by exerting an inward
force to balance the repulsion from within. These types of {\it
grav}itationally {\it va}cuum {\it stars} were termed as {\it
gravastars}. Thereafter several scientists have been studied these
models under different viewpoints and have opened up a new field of
research as an alternative to {\it Black
Holes}~\cite{Visser2004,Cattoen2005,Carter2005,Bilic2006,Lobo2006,DeBenedictis2006,Lobo2007,Horvat2007,Cecilia2007,Rocha2008,Horvat2008,Nandi2009,Turimov2009}.

Very recently, a charged $(3+1)$-dimensional gravastar admitting
conformal motion has proposed by some of our collaborators~\cite{Usmani2011}
in the framework of Mazur and Mottola model~\cite{Mazur2001,Mazur2004}. In this
work the authors provide an alternative to static black holes.
However, energy density here is found to diverge in the interior
region of the gravastar. This actually scales like an inverse
second power of its radius and unfortunatelly makes the model singular
at $r=0$. However, interestingly in one of the solutions it is shown that
the total gravitational mass vanishes for vanishing charge and turns the
total gravitational mass into an electromagnetic mass under certain conditions.
An extention on charged gravastar of Usmani et al.~\cite{Usmani2011}
can be found in the work of Bhar~\cite{Bhar2014} admitting conformal
motion with higher dimensional space-time.

In the present study we generalize the four-dimensional work on gravastar
by Usmani et al.~\cite{Usmani2011} to the higher dimensional space-time, however
without admitting conformal motion. Our main motivation here is to
construct gravastars in the Einstein-Maxwell geometry and see the
higher dimensional effects, if any. Therefore this investigation
is also extension of the work of Bhar~\cite{Bhar2014} without
admitting conformal motion and that of Rahaman et al.~\cite{Rahaman2015}
with charge where originally higher dimensional gravastar has been studied.
A detailed discussion on higher dimension and its applications in various fields
of astrophysics as well as cosmology has been provided in Ref.~\cite{Rahaman2015}.

The plan of the present investigation is as follows: In Sec. 2
the Einstein-Maxwell space-time geometry has been provided as the
background of the study whereas in Sec. 3 we discuss the Interior
space-time, Exterior space-time and Thin shell cases of the
gravastars with their respective solutions. The related
junction conditions are provided in Sec. 4. We explore physical
features of the models, viz. proper length, energy condition, entropy, mass and
equation of state in Sec. 5. At the end in Sec. 6 we
provide some critically discussed concluding remarks.

\section{The Einstein-Maxwell space-time geometry}

For higher dimensional gravastar, we assume a $D$-dimensional
spacetime with the typical mathematical structure $R^1 X S^1 X S^d~(d = D - 2)$, where
$S^1$ is the range of the radial coordinate $r$ and $R^1$ is the
time axis. Let us therefore consider a static spherically
symmetric metric in $D = d + 2$ dimension as
\begin{equation}
ds^2 = -e^{\nu}dt^2 + e^{\lambda}dr^2+r^2 d\Omega_d ^2,\label{eq1}
\end{equation}
where $ d\Omega_d ^2$ is a linear element on a
$d$-dimensional unit sphere, parametrized by the angles $\phi_1,
\phi_2,...,\phi_d$, as follows: $d\Omega_d ^2= d\phi_d^2 + \sin_2 \phi_d [d\phi_{d-1}^2 + \sin_2
\phi_{d-1}\{d\phi_{d-2}^2 + ...+ \sin_2 \phi_3(d\phi_2^2 +
\sin_2 \phi_2  d\phi_1^2)...\}] $.

Now, the Hilbert action coupled to matter and electromagnetic field can be provided as
\begin{equation}
I = \int d^D x \sqrt{-g } \left[\frac{R_D}{16 \pi G_D} + ({L}_{m} + {F}_{ik}{F}^{ik})\right],\label{eq2}
\end{equation}
where $L_m$ is the matter part of the Lagrangian and  $F_{ij}$ is the electromagnetic field tensor which is related to the
electromagnetic potentials through the relation $ F_{ij} = A_{i,j} - A_{j,i}.$

In the above Eq. (\ref{eq2}) the term $R_D$ is the curvature scalar
in $D$-dimensional spacetime whereas $G_D$ is the $D$-dimensional
Newtonian constant and $L_{m}$ is the Lagrangian for matter-energy distribution.

The Einstein-Maxwell field equations now can be written as
\begin{equation}
G^D_{ij} = - 8 \pi G_D [T_{ij}^m+T_{ij}^{em}],\label{eq3}
\end{equation}
where $G^D_{ij}$ is the Einstein tensor in $D$-dimensional
spacetime, $T_{ij}^m$ and $T_{ij}^{em}$ are the matter-energy and electromagnetic
tensors respectively.

We assume that the interior of the star is filled up with perfect fluid and
therefore the matter-energy tensors can be considered in the following form
\begin{equation}
T_{ij}^m = (\rho + p ) u_i u_j + p  g_{ij}, \label{eq4}
\end{equation}
where $\rho$ is the energy density, $p$ is the isotropic
pressure and  $u^{i}$  (with $u_{i}u^{i}=1$) is the $D$-velocity of the fluid under consideration.

On the other hand, the electromagnetic tensors can be provided as
\begin{equation}
T_{ij}^{em} = -\frac{1}{4\pi {G}_D }\left[F_{jk}F_i^k - \frac{1}{4} g_{ij}F_{kl} F^{kl}\right]. \label{eq5}
\end{equation}

The corresponding Maxwell electromagnetic field equations are
\begin{equation}
{[{(- g)}^{1/2} F^{ij}],}_{j} = 4\pi J^{i}{(- g)}^{1/2},\label{eq6}
\end{equation}

\begin{equation}
F_{[ij,k]} = 0,\label{max2},\label{eq7}
\end{equation}
where $J^{i}$ is the current four-vector satisfying $J^{i} = \sigma u^{i}$, the parameter $\sigma$ being the charge density.

Hence the Einstein-Maxwell field equation (\ref{eq3}), for the metric (\ref{eq1}) along with
the energy-momentum tensors, Eqs.~(\ref{eq4}) - (\ref{max2}), can be provided in the following explicit forms
\begin{equation}
-e^{-\lambda} \left[\frac{d(d-1)}{2r^2} - \frac{d \lambda'}{2r}
\right] + \frac{d(d-1)}{2r^2} = 8\pi G_D~ \rho + E^2, \label{eq8}
\end{equation}

\begin{equation}
e^{-\lambda} \left[\frac{d(d-1)}{2r^2} + \frac{d \nu'}{2r} \right]
- \frac{d(d-1)}{2r^2} = 8\pi G_D ~p -E^2, \label{eq9}
\end{equation}

\begin{eqnarray}
\frac{e^{-\lambda}}{2} \left[ \nu'' -\frac{\lambda'\nu'}{2}
+\frac{ {\nu'}^2}{2} -\frac{(d-1)(\lambda'-\nu')}{r} + ~~~~~~~~~~~~~~~~~~~~~\right. \nonumber \\ \left.
\frac{(d-1)(d-2) }{r^2} \right] - \frac{(d-1)(d-2)
}{2r^2} = 8\pi G_D~p +E^2,~~~~~~~~~~~~\nonumber\\ \label{eq10}
\end{eqnarray}
where $E$ is the electric field. Here the symbol `$\prime$' denotes
differentiation with respect to the radial parameter $r$ and $c = 1$ (in geometrical unit).

Therefore, the energy conservation equation in the $D$-dimensions is given by
\begin{equation}
\frac{1}{2} \left(\rho + p\right)\nu' + p' =\frac{1}{4\pi{G}_D
r^d}(r^dE^2)', \label{eq11}
\end{equation}
with the electric field $E$ as follows
\begin{equation}
(r^dE)'=\frac{2 \pi^{\frac{d+1}{2}}}{\Gamma(\frac{d+1}{2})} r^d\sigma (r) e^{\lambda/2}.\label{eq12}
\end{equation}

In traditional sense, the term $\sigma e^{\lambda/2}$ appearing in the right hand side of Eq. (12), is known as  the volume charge density. The assumption $\sigma e^{\lambda/2}= \sigma_0 r^m$, can consistently be understood as the higher dimensional
volume charge density being polynomial function of $r$ where the constant $\sigma_{0}$ is the central charge density.

Now from the above Eq. (\ref{eq12}) by assuming $\sigma e^{\lambda/2}= \sigma_0 r^m$, we obtain the explicit form of the electric field as given by
\begin{equation}
E= \frac{q}{r^d}= \frac{2 \pi^{\frac{d+1}{2}}\sigma_{0}}{\Gamma(\frac{d+1}{2})} \frac{r^{m+1}}{(d+m+1)}  =
A r^{m+1},\label{eq13}
\end{equation}
where $A=\frac{2 \pi^{\frac{d+1}{2}}\sigma_{0}}{\Gamma(\frac{d+1}{2})(d+m+1)}$.

\section{The gravastar models}

\subsection{Interior space-time}

Following \cite{Mazur2001} we assume that the
EOS for the interior region has the form
\begin{equation}
p = -\rho. \label{eq14}
\end{equation}

The above EOS is known in the literature as a `false vacuum',
`degenerate vacuum', or
`$\rho$-vacuum'~\cite{Davies1984,Blome1984,Hogan1984,Kaiser1984}
which represents a repulsive pressure, an agent
responsible for the accelerating phase of the Universe, and is
termed as the $\Lambda$-dark
energy~\cite{Riess1998,Perlmutter1999,Ray2007b,Usmani2008,Frieman2008}.
It is argued by \cite{Usmani2011} that a charged
gravastar seems to be connected to the dark
star~\cite{Lobo2008,Chan2009a,Chan2009b}.

The above EOS along with Eq. (\ref{eq11}) readily provides
\begin{equation}
p=-\rho=k_1 {r^{2(m+1)}}+k_2. \label{eq15}
\end{equation}
where $k_1=\frac{A^2(2m+d+2)}{4\pi {G}_D (2m+2)}$ and $k_2$ is an integration constant. However, if we put $r=0$  in Eq. (\ref{eq15}), then it easily assigns the value of the integration constant $k_2=p_c=-\rho_c$. In general there is no sufficient argument to take pressure and density to be zero at the junction surface. Actually, in the thin shell limit the pressure and density are step functions at the junction surface~\cite{MM15}. However, for the sake of brevity and convenience, if one considers boundary condition on the spherical surface that at $r=R$ the pressure and density in Eq. (\ref{eq15}) vanish, then it yields $k_2=-k_1R^{2(m+1)}=- \rho_c$, i.e. through $k_1$ now $k_2$ is also dependent on the charge density $\sigma_0$. This simply provides an expression for the central density,  $\rho_c = k_1R^{2(m+1)}$ and thus in the present case we are treating with a spherical stellar system with a constant central density or equivalently pressure which is acting outwardly.

If one agrees with the above argument, then one can trust on the relation between $k_1$  and $k_2$, and also to the charge density $σ_0$ as a consequent. This, therefore, makes someone to speculate about the ``electromagnetic mass model (EMM).'' It is to be noted that the spherically symmetric system being static the magnetic counterpart will not act here at all and hence the matter stress tensor is not completely electromagnetic in origin rather the electric part is associated only. However, even in this case one can borrow the phrase ``electromagnetic mass model (EMM)'' because of the fact that Lorentz~\cite{Lorentz1904} himself termed it like that and we are in the present situation coining the phrase due to historical reasons only.

Using Eq. (\ref{eq15}) in the field equation (\ref{eq8}), we get the expression of the metric potential $\lambda$ given by
\begin{eqnarray}
e^{-\lambda} = 1- \frac{2A^2}{d(d+2m+3)}r^{2(m+2)} +\frac{16\pi G_Dk_1}{d(d+2m+3)}r^{2(m+2)}~~~~~~~~~\nonumber\\ +\frac{16\pi G_Dk_2}{d(d+1)}r^2 +k_3r^{1-d},\label{eq16}
\end{eqnarray}
where $k_3$ is an integration constant. Since $d>2$ for dimension higher
 than four and the solution is regular at $r=0$, so one can demand for $k_3=0$. Thus, Eq. (\ref{eq16}) essentially reduces to
\begin{eqnarray}
e^{-\lambda} = 1- \frac{2A^2}{d(d+2m+3)}r^{2(m+2)} +\frac{16\pi G_Dk_1}{d(d+2m+3)}r^{2(m+2)}+\frac{16\pi G_Dk_2}{d(d+1)}r^2 . \label{eq17}
\end{eqnarray}

Again, using Eqs. (\ref{eq8}), (\ref{eq9}) and (\ref{eq15}) one may obtain the following unique relation
\begin{equation}
\ln k =\lambda + \nu,\label{eq18}
\end{equation}
where $k$ is an integration constant.

Thus we have the interior solutions for the metric
potentials $\lambda$ and $\nu$ as follows
\begin{eqnarray}
e^\nu=ke^{-\lambda} =k\left[ 1- \frac{2A^2}{d(d+2m+3)}r^{2(m+2)} +\frac{16\pi G_Dk_1}{d(d+2m+3)}r^{2(m+2)}\right.~~~~~~~~~\nonumber\\ \left.+\frac{16\pi G_Dk_2}{d(d+1)}r^2 \right],~~~~\nonumber\\\label{eq19}
\end{eqnarray}

The active gravitational mass $M(r)$ in higher dimensions, can be
now calculated as
\begin{eqnarray}
M(R) = \int_0^{{r_1=R}}~ \left[  \frac{2 \pi^{\frac{d+1}{2}}
}{\Gamma \left(\frac{d+1}{2}\right)}\right]r^d \left[\rho+\frac{E^2}{8\pi}\right] dr~~~~~~~~~~~~~ \nonumber\\= \frac{2 \pi^{\frac{d+1}{2}} }{\Gamma\left(\frac{d+1}{2}\right)} \left[\frac{A^2}{8\pi}\frac{R^{2m+d+3}}{2m+d+3}-\frac{k_1R^{2m+d+3}}{2m+d+3}-\frac{k_2R^{d+1}}{d+1}\right].\label{eq20}
\end{eqnarray}

This is the usual gravitating mass for a $d$-dimensional sphere of
radius $R$ and energy density $\rho$. The space-time metric here
turns out to be free from any central singularity. Another interesting
point one can easily observe that
the density, pressure and mass  (Eqs. (\ref{eq15}) and (\ref{eq20})) do vanish
and via the metric potentials (Eqs. (\ref{eq19})) space-time becomes flat for vanishing
charge density $\sigma_0$. Therefore, the interior solutions
provide electromagnetic mass (EMM) model
~\cite{Lorentz1904,Wheeler1962,Feynman1964,Wilczek1999,Florides1962,Cooperstock1978,Tiwari1984,Gautreau1985,Gron1985,Leon1987,Ray2004,Ray2006,Ray2008,Usmani2011,Rahaman2012}.
This result suggests that the interior de Sitter vacuum of a
charged gravastar is essentially an electromagnetic mass
that must generate the gravitational mass~\cite{Usmani2011}. A
detailed discussion on this EMM model can be obtained in
Ref.~\cite{Rahaman2015}.

\subsection{Exterior space-time}
The exterior region defined by $p=\rho=0$ in higher dimensions
is nothing but a generalization of Reissner-Nordstr{\"o}m solution.
Therefore, following \cite{Tangherlini1963} the Reissner-Nordstr{\"o}m metric can be obtained as
\begin{eqnarray}
ds ^2 = - \left(1 - \frac{\mu}{
r^{d-1}}+\frac{q^2}{r^{2(d-1)}}\right) dt^2~~~~~~~~~~~~~~~~ \nonumber \\+ \left(1 -
\frac{\mu}{ r^{d-1}}+\frac{q^2}{r^{2(d-1)}}\right)^{-1} dr^2 + d
\Omega_d ^2.\label{eq21}
\end{eqnarray}

Here $\mu=16\pi G_D M/\Omega_d$ is the constant of integration
with $M$, the  mass of the black hole and ${\Omega}_d$, the area
of a unit $d$-sphere as ${\Omega}_d = 2
\pi^{(\frac{d+1}{2})}/\Gamma(\frac{d+1}{2})$.

\subsection{Intermediate thin shell}
Here we assume that the thin shell contains ultra-relativistic
fluid of soft quanta and obeys the EOS
\begin{equation}
p =\rho. \label{eq22}
\end{equation}

This relation represents essentially a stiff fluid model as
envisioned by \cite{Zeldovich1972} in connection to
cold baryonic universe and have been considered by several authors
for various situations in cosmology~\cite{Carr1975,Madsen1992} as
well as in astrophysics~\cite{Buchert2001,Braje2002,Linares2004}.

It is difficult to obtain a general solution of the field
equations in the non-vacuum region, i.e. within the shell. We try
to find an analytic solution  within the thin shell limit,
 $0< (e^{-\lambda}\equiv h) <<1$.  As an advantage of it, we may set $h$ to
be zero to the leading order. Under this approximation, the field
Eqs. (\ref{eq8}) - (\ref{eq10}), with the above EOS, may be recast
in the following form
\begin{equation}
 \frac{h'}{2 r}  = \frac{(d-1)}{r^2}-\frac{2E^2}{d},\label{eq23}
\end{equation}

\begin{equation}
\frac{\nu'  h'}{4} +\frac{(d-1)h'}{2r}= - \frac{(d-1)}{r^2}
+2E^2\label{eq24}.
\end{equation}

Integration of Eq. (20) immediately yields
\begin{equation}
h = 2 (d-1) \ln r - \frac{2A^2 r^{2(m+2)}}{d(m+2)} + c_{1},\label{eq25}
\end{equation}
 where $c_{1}$ is an integration constant. The other metric potential can be found as
 \begin{equation}
 e^{\nu}=\frac{c_2}{r^{2(2d-1)}}\left(\frac{r^{2m+4}}{d(d-1)-2A^2r^{2m+4}}\right)^{\frac{d-1}{m+2}},\label{eq26}
 \end{equation}
where $c_2$ is an integration constant.

%%%%%%%%%%%%%%%%%%%%%%%%%%%%%%%%%%%%%%%%
\begin{figure*}[thbp]
\centering
\includegraphics[width=0.5\textwidth]{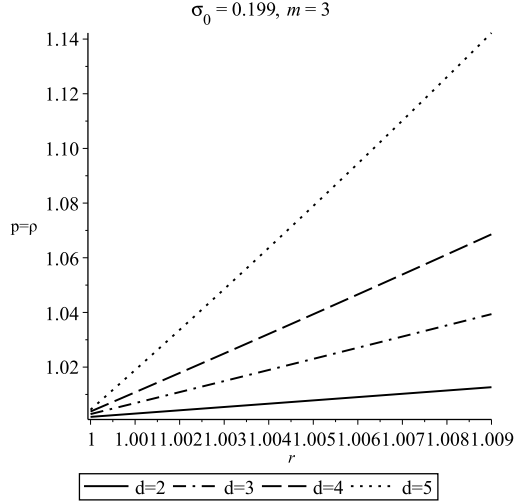}
\caption{Variation of the pressure and density of the ultra
relativistic matter in the shell against $r$ for different
dimensions where the specific legends used are shown in the
respective plots}
\end{figure*}
%%%%%%%%%%%%%%%%%%%%%%%%%%%%%%%%%%%%%%%%%

Now in order to calculate the pressure as well as the matter
density within the thin shell, using the Eq. (8) and the EOS Eq. (19),
 one may obtain
\begin{equation}
p = \rho = B \epsilon r^{2m+1} + c_3 e^{-\nu}, \label{eq27}
\end{equation}
where $B=\frac{A^2(2m+d+2)}{4\pi{G}_D }$ and $c_3$ is an integration constant. From Eq. (27) it can be
observed that the pressure and matter density is depends on the
central charge density and the thickness of the shell. The variation
of the matter density over the shell is shown in Fig. 1 for different
dimensions which shows that the matter density is increasing
from the interior boundary to the exterior boundary.

\section{Junction Condition}
As mentioned earlier, in the  gravastar configuration there are three regions:
interior, shell and exterior region. Interior and exterior
regions join at the junction interface of the shell. Though at the
shell the metric coefficients are continuous but to see whether their
derivatives are also continuous there we follow the Darmois-Israel
condition~\cite{Darmois1927,Israel1966} to
calculate the surface stresses at the junction interface. Therefore
following the Lancozs equations for the intrinsic surface stress energy
tensor $S_{ij}$, we can determine the surface energy density $(\Sigma)$ and surface
pressure $p_{\theta_{1}}=p_{\theta_{2}}=...=p_{\theta_{d-1}}=p_t$
as
\begin{eqnarray}
\Sigma=-\frac{d}{4\pi R}\left[\sqrt{1-\frac{\mu}{R^{d-1}}+\frac{q^{2}}{R^{2(d-1)}}}- ~~~~~~~~~~~~~~~~~~~~~~~\right. \nonumber\\
\left.\sqrt { 1+\frac{16\pi G_Dk_1R^{2(m+2)}}{d(d+2m+3)}-\frac{2A^2R^{2(m+2)}}{d(d+2m+3)}
+\frac{16\pi G_Dk_2R^2}{d(d+1)}}\right],\label{eq28}
\end{eqnarray}

\begin{eqnarray}
p_t=\frac{1}{8\pi
R}\left[\frac{2(d-1)-\frac{(d-1)\mu}{R^{d-1}}}{\sqrt{1-\frac{\mu}{R^{d-1}}+\frac{q^{2}}{R^{2(d-1)}}}} -  ~~~~~~~~~~~~~~~~~~~~~~~~~~~~~~~~ \right. \nonumber\\
\left.\frac{ {2(d-1)+\frac{32\pi G_Dk_1(d+m+1)R^{2(m+2)}}{d(d+2m+3)}-\frac{4A^2(d+m+1)R^{2(m+2)}}{d(d+2m+3)}
+\frac{32\pi G_Dk_2 R^2}{d+1}}}{\sqrt { 1+\frac{16\pi G_Dk_1R^{2(m+2)}}{d(d+2m+3)}-\frac{2A^2R^{2(m+2)}}{d(d+2m+3)}
+\frac{16\pi G_Dk_2R^2}{d(d+1)}}}  \right].\label{eq29}
\end{eqnarray}

\section{Physical features of the models}

\subsection{Proper length}
We consider that matter shell is situated at the surface $ r = R$,
describing the phase boundary of region I. The thickness of the
shell ($\epsilon << 1$) is assumed to be very small. Thus the
region III joins at the surface $ r = R+\epsilon $.

Now, we calculate the proper thickness between two interfaces i.e.
of the shell as
\begin{eqnarray}
\ell = \int _{R}^{R+\epsilon} \sqrt{e^{\lambda} } dr= \int_{R}^{R+\epsilon} \frac{dr}{\sqrt{[c_{1}+ 2 (d-1) \ln r - \frac{2A^2 r^{2(m+2)}}{d(m+2)}]}}=\int_{R}^{R+\epsilon}\frac{dr}{f(r)}.\label{eq30}
\end{eqnarray}
where $f(r)=\sqrt{[c_{1}+ 2 (d-1) \ln r - \frac{2A^2 r^{2(m+2)}}{d(m+2)}]}$.

To solve the above equation, let us consider that $\frac{df(r)}{dr}=\frac{1}{f(r)}$, so that we get
\begin{eqnarray}
\ell=[f(R+\epsilon)-f(R)]\approx\epsilon\frac{df}{dr}\approx \frac{\epsilon}{\sqrt{[c_{1}+ 2 (d-1) \ln r - \frac{2A^2 r^{2(m+2)}}{d(m+2)}]}}. \label{eq31}
\end{eqnarray}

As the value of $\epsilon$ is very small so the higher order
terms of $\epsilon$ can be neglected. The variation of proper
length for different dimensions with polynomial index is shown in
Fig. 2.

%%%%%%%%%%%%%%%%%%%%%%%%%%%%%%%%%%%%%%%%
\begin{figure*}[thbp]
\centering
\includegraphics[width=0.5\textwidth]{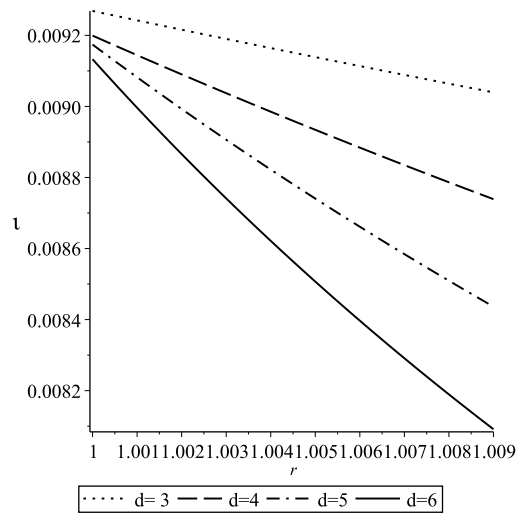}
\caption{Variation of the length against $r$ for different dimensions
where the specific legends used are shown in the respective plots}
\end{figure*}
%%%%%%%%%%%%%%%%%%%%%%%%%%%%%%%%%%%%%%%%

\subsection{Energy}
The energy $E$ within the shell can be calculated as
\begin{eqnarray}
E  =   \int _{R}^{R+\epsilon}\left[\frac{2
\pi^{\frac{d+1}{2}} } {\Gamma
\left(\frac{d+1}{2}\right)}\right]r^d \left [\rho +\frac
{E^2}{8\pi}\right]dr
\approx\left[  \frac{2 \pi^{\frac{d+1}{2}}
}{\Gamma\left(\frac{d+1}{2}\right)}\right]  \left[\frac{\epsilon A^2}{8\pi}R^{2(m+1)+d} \right]. \label{eq32}
 \end{eqnarray}

%%%%%%%%%%%%%%%%%%%%%%%%%%%%%%%%%%%%%%%%
\begin{figure*}[thbp]
\centering
\includegraphics[width=0.5\textwidth]{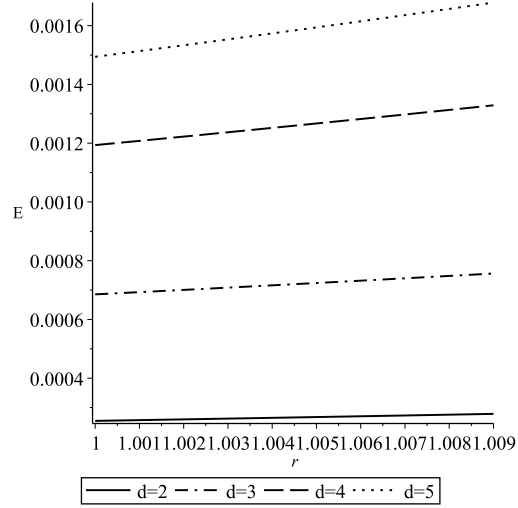}
\caption{Variation of energy with $r$ for different dimensions
where the specific legends used are shown in the respective plots}
\end{figure*}
%%%%%%%%%%%%%%%%%%%%%%%%%%%%%%%%%%%%%%%%

We have considered the energy ${E}$ within the shell up to the first order in
$\epsilon$. The thickness $\epsilon$ of the
shell being very small ($\epsilon << 1$), we expand it
binomially about $R$ and consider first order of $\epsilon$ only.
It can be observed that the energy within the shell (i) is
proportional to $\epsilon$ in first order of thickness, and
(ii) depends on the dimension $d$ of the spacetime. Fig. 3
shows energy profile of charged gravastar in higher dimension.

\subsection{Entropy}
Following Mazur and Mottola prescription~\cite{Mazur2001} the entropy can be given as
\begin{equation}
 S =  \int _{R}^{R+\epsilon} \left[  \frac{2
\pi^{\frac{d+1}{2}} }{\Gamma \left(\frac{d+1}{2}\right)}\right]r^d
s(r)   \sqrt{e^{\lambda}}dr,\label{eq33}
\end{equation}
where $s(r)$ stands for the entropy density of the local
temperature $T(r)$, which is written as
\begin{equation}
s(r) =  \frac{\xi^2k_B^2T(r)}{4\pi\hbar^2 } =
\xi\left(\frac{k_B}{\hbar}\right)\sqrt{\frac{p}{2 \pi
}},\label{eq34}
\end{equation}
where $\xi$ is a dimensionless constant.

Now the entropy of the fluid within the shell can be given by
\begin{eqnarray}
S \approx \left[\frac{2 \pi^{\frac{d+1}{2}}}{\Gamma \left(\frac{d+1}{2}\right)}\right]\frac{\xi k_B\epsilon r^d}{\hbar\sqrt{2\pi}} \sqrt{\frac{B\epsilon r^{2m+1}+c_3 e^{-\nu}}{c_1+2(d-1)lnr -\frac{2A^2r^{2(m+2)}}{d(m+2)}}}.\label{eq35}
\end{eqnarray}

Following Ref.~\cite{Usmani2011} it can be shown that (i)
the thickness of the shell is negligibly small compared to
its position from the center of the gravastar (i.e., if $\epsilon
<< R$), and (ii) the entropy depends on the thickness of the shell.

\subsection{Mass} Now, it is easy to find the surface mass $m_s$ of the thin
shell from the equation
\begin{eqnarray}
m_s=\frac{2\pi^{\frac{d+1}{2}}}{\Gamma(\frac{d+1}{2})}R^d\Sigma=\frac{2\pi^{\frac{d+1}{2}}}{\Gamma(\frac{d+1}{2})}\frac{dR^{d-1}}{4\pi} \Bigg[X -\sqrt{1-\frac{\mu}{R^{d-1}}+\frac{q^{2}}{R^{2(d-1)}}}\Bigg], \label{eq36}
\end{eqnarray}
where $X=\sqrt { 1+\frac{16\pi G_Dk_1R^{2(m+2)}}{d(d+2m+3)}-\frac{2A^2R^{2(m+2)}}{d(d+2m+3)}
+\frac{16\pi G_Dk_2R^2}{d(d+1)}}.$

Using Eq. (36) we can determine the total mass of the
gravastar in terms of the mass of the thin shell as
\begin{eqnarray}
\mu=\frac{q^2}{R^{d-1}}-\left[
\frac{16\pi G_Dk_1R^{2m+d+3}}{d(d+2m+3)}-\frac{2A^2R^{2m+d+3}}{d(d+2m+3)} +\frac{16\pi G_Dk_2R^{d+1}}{d(d+1)}\right. \nonumber\\
\left.+\frac{m_s^2 }{F^2 R^{d-1} } -\frac{2m_s}{F}X \right],\nonumber\\ \label{eq37}
\end{eqnarray}
where $F=\frac{2\pi^{\frac{d+1}{2}}}{\Gamma(\frac{d+1}{2})}\left[\frac{d}{4\pi}\right]$.

\subsection{Equation of State}
Let us assume
$p_{\theta_{1}}=p_{\theta_{2}}=p_{\theta_{3}}=...=p_t=-\mathcal{P}$,
here $\mathcal{P}$ is the surface tension as acts on the fluid of
the gravastar.

Then Eqs. (28) and (29) yield
\begin{equation}
\mathcal{P}=\omega(R)\Sigma.\label{eq38}
\end{equation}
Thus the Equation of State parameter $\omega$ can be found as
\begin{eqnarray}
\omega(R)=\frac{-1}{2d}\left[\frac{\frac{2(d-1)-(d-1)\frac{\mu}{R^{d-1}}}{\sqrt{1-\frac{\mu}{R^{d-1}}+\frac{q^{2}}{R^{2(d-1)}}}}+
\frac{Y}{X} }{\sqrt{1-\frac{\mu}{R^{d-1}}+\frac{q^{2}}{R^{2(d-1)}}}-X}\right],\label{eq39}
\end{eqnarray}
where $Y= {2(d-1)+\frac{32\pi G_Dk_1(d+m+1)R^{2(m+2)}}{d(d+2m+3)}-\frac{4A^2(d+m+1)R^{2(m+2)}}{d(d+2m+3)}
+\frac{32\pi G_Dk_2 R^2}{d+1}}$.

\section{Discussions and Conclusions}
In the present study we have explored some possibilities to find
out a new model of gravastars in contrast to the Mazur-Mottola
type model of four-dimensional and neutral gravastar~\cite{Mazur2001,Mazur2004},
specifically seeking its generalization to: (i) the extended
$D$-dimensional spacetime, and (ii) the Einstein-Maxwell geometry.
Under these two considerations we have found out a class of solutions
and hence some interesting results which can be observed as
an alternative to $D$-dimensional versions of the
Reissner-Nordstr{\"o}m-Tangherlini black hole.

Some of the key physical features of the model are as follows:

(i) We have found out all the physical parameters e.g. metric
potentials, thickness of the thin shell, energy, entropy etc. and
our result matches to the result of~\cite{Usmani2011} for $d = 2$ and $m=0$ i.e. for
$4$-dimensional space-time without any polynomial index. All the
plots (Figs. 1 - 3) related to these parameters also suggest
validity of physical requirements.

(ii) It is interesting to note that all the solutions are regular
at the center $r=0$ and positive inside the interior region of the
gravastar. Specifically, if we put $r=0$ in Eq. (\ref{eq15}) then via
the integration constant $k_2=p_c=-\rho_c$ we get a spherical system
with a constant central density or pressure which being incompressible
makes the gravastar free from singularity as well as stable.

(iii) We observe (from Eqs. (\ref{eq15}) and (\ref{eq20}))
that the density, pressure and mass like all the physical
parameters do vanish and also the space-time becomes flat for vanishing
charge density $\sigma_0$. Therefore, the interior solutions
provides electromagnetic mass (EMM) model
~\cite{Lorentz1904,Wheeler1962,Feynman1964,Wilczek1999,Florides1962,Cooperstock1978,Tiwari1984,Gautreau1985,Gron1985,Leon1987,Ray2004,Ray2006,Ray2008,Usmani2011,Rahaman2012}.
This result suggests that unlike the work of Usmani et al.~\cite{Usmani2011}
and Rahaman et al.~\cite{Rahaman2012} the interior de Sitter vacuum of a
charged gravastar is always an electromagnetic mass which must generate
the gravitational mass rather in the present case a constant central pressure
acts as repulsive force which prevents the system to undergo the fatal singularity .

As a final comment, we would like to put an important note here
regarding overall observational results of the present
investigation on gravastar with higher dimensional space-time. As
a sample study a comparison between Figs. 1 - 3 shows that
there are lots of quantitative change in the profiles of the physical
parameters and as one goes on towards higher dimentions
much appreciable results can be observed. All these observational
surveys are probable indication of applicability of higher
dimensional approach for construction of a gravastar with or
without charge from an ordinary $4$-dimensional seed. In connection
to this concluding remark we note that in the work of Bhar et al.~\cite{Bhar2015},
where they have performed an investigation on the possibility of
higher dimensional compact stars, the results are in favour of our
present study.

\section*{Acknowledgments}
FR and SR wish to thank the authorities of the Inter-University
Centre for Astronomy and Astrophysics, Pune, India for providing
the Visiting Associateship under which a part of this work was
carried out.  SR is also thankful to The Institute of
Mathematical Sciences (IMSc), Chennai, India for providing
Visiting Associateship. FR is thankful to the DST, Govt. of India for
financial support under PURSE programme (EMR/2016/000193). We are really grateful
to the referee for several critical remarks and fruitful suggestions
which has led the development of the paper substantially.

\end{document}